\documentstyle[psfig]{mn}

\newif\ifAMStwofonts

\ifoldfss
  \ifCUPmtlplainloaded \else
    \NewTextAlphabet{textbfit} {cmbxti10} {}
    \NewTextAlphabet{textbfss} {cmssbx10} {}
    \NewMathAlphabet{mathbfit} {cmbxti10} {} 
    \NewMathAlphabet{mathbfss} {cmssbx10} {} 
  \fi
  \ifAMStwofonts
    \ifCUPmtlplainloaded \else
      \NewSymbolFont{upmath} {eurm10}
      \NewSymbolFont{AMSa} {msam10}
      \NewMathSymbol{\upi}     {0}{upmath}{19}
      \NewMathSymbol{\umu}     {0}{upmath}{16}
      \NewMathSymbol{\upartial}{0}{upmath}{40}
      \NewMathSymbol{\leqslant}{3}{AMSa}{36}
      \NewMathSymbol{\geqslant}{3}{AMSa}{3E}

      \let\geq=\geqslant 
    \fi
  \fi
\fi 

\ifnfssone
  \newmathalphabet{\mathit}
  \addtoversion{normal}{\mathit}{cmr}{m}{it}
  \addtoversion{bold}{\mathit}{cmr}{bx}{it}
  \newmathalphabet{\mathbfit} 
  \addtoversion{normal}{\mathbfit}{cmr}{bx}{it}
  \addtoversion{bold}{\mathbfit}{cmr}{bx}{it}
  \newmathalphabet{\mathbfss} 
  \addtoversion{normal}{\mathbfss}{cmss}{bx}{n}
  \addtoversion{bold}{\mathbfss}{cmss}{bx}{n}
  \ifAMStwofonts
    \ifCUPmtlplainloaded \else
      \UseAMStwoboldmath
      \makeatletter
      \new@mathgroup\upmath@group
      \define@mathgroup\mv@normal\upmath@group{eur}{m}{n}
      \define@mathgroup\mv@bold\upmath@group{eur}{b}{n}
      \edef\UPM{\hexnumber\upmath@group}
      \new@mathgroup\amsa@group
      \define@mathgroup\mv@normal\amsa@group{msa}{m}{n}
      \define@mathgroup\mv@bold\amsa@group{msa}{m}{n}
      \edef\AMSa{\hexnumber\amsa@group}
      \makeatother
      \mathchardef\upi="0\UPM19
      \mathchardef\umu="0\UPM16
      \mathchardef\upartial="0\UPM40
      \mathchardef\leqslant="3\AMSa36
      \mathchardef\geqslant="3\AMSa3E

      \let\geq=\geqslant 
    \fi
  \fi
\fi 

\ifnfsstwo
  \DeclareMathAlphabet{\mathbfit}{OT1}{cmr}{bx}{it}
  \SetMathAlphabet\mathbfit{bold}{OT1}{cmr}{bx}{it}
  \DeclareMathAlphabet{\mathbfss}{OT1}{cmss}{bx}{n}
  \SetMathAlphabet\mathbfss{bold}{OT1}{cmss}{bx}{n}
  \ifAMStwofonts
    \ifCUPmtlplainloaded \else
      \DeclareSymbolFont{UPM}{U}{eur}{m}{n}
      \SetSymbolFont{UPM}{bold}{U}{eur}{b}{n}
      \DeclareSymbolFont{AMSa}{U}{msa}{m}{n}
      \DeclareMathSymbol{\upi}{0}{UPM}{"19}
      \DeclareMathSymbol{\umu}{0}{UPM}{"16}
      \DeclareMathSymbol{\upartial}{0}{UPM}{"40}
      \DeclareMathSymbol{\leqslant}{3}{AMSa}{"36}
      \DeclareMathSymbol{\geqslant}{3}{AMSa}{"3E}

      \let\geq=\geqslant 
    \fi
  \fi
\fi 

\ifCUPmtlplainloaded \else
  \ifAMStwofonts \else 
    \def\upi{\pi}
    \def\umu{\mu}
    \def\upartial{\partial}
  \fi
\fi

\def\TCC{\mbox{$C_{r_{\rm e}}$}}

\title[On the estimation of galaxy structural parameters]
{On the estimation of galaxy structural parameters: the S\'ersic Model}
\author[I. Trujillo et al.]
       {I. Trujillo$^{1}$\thanks{itc@ll.iac.es}, Alister W. Graham$^{1}$ and N. Caon$^{1}$ 
        \\
        $^{1}$ Instituto de Astrof\'{\i}sica de Canarias,  E-38205 La Laguna, 
	Tenerife, Spain\\}
\date{Accepted 0000 December 00.
      Received 0000 December 00;
      in original form 0000 October 00}

\pagerange{\pageref{firstpage}--\pageref{lastpage}}
\pubyear{2000}

\begin{document}

\maketitle

\label{firstpage}

\begin{abstract} 

This paper addresses some questions which have arisen from the use of the
S\'ersic $r^{1/n}$ law in modelling the luminosity profiles of early type
galaxies. The first issue deals with the trend between the half-light
radius and the structural parameter $n$.
We show that the correlation between these two parameters is not only real,
but is a natural consequence from the previous relations found to exist between
the model-independent parameters: total luminosity, effective radius and
effective surface brightness.  We also define a new galaxy concentration index
which is largely independent of the image exposure depth, and monotonically
related with $n$.
The second question concerns the curious coincidence between the form of
the Fundamental Plane and the coupling between $<$$I$$>$$_e$ and $r_e$
when modelling a light profile.  We explain, through a mathematical
analysis of the S\'ersic law, why the quantity
$r_e$$<$$I$$>$$_e^{0.7}$ appears almost constant for an individual galaxy, 
regardless of the value of $n$ (over a large range) adopted in the fit 
to the light profile.
Consequently, Fundamental Planes of the form
$r_e$$<$$I$$>$$_e^{0.7}$$\propto$$\sigma_0^x$ (for any $x$, and where
$\sigma_0$ is the central galaxy velocity dispersion) are insensitive
to galaxy structure.
Finally, we address the problematic issue of the use of model-dependent
galaxy light profile parameters versus model-independent quantities for
the half-light radii, mean surface brightness and total galaxy magnitude.
The former implicitly assume that the light profile model can be extrapolated
to infinity, while the latter quantities, in general, are derived from a
signal-to-noise truncated profile.  We quantify (mathematically) how these
parameters change as one reduces the outer radius of an $r^{1/n}$ profile,
and reveal how these can vary substantially when $n$$\geq$4.

\end{abstract}


\section{Introduction}

The parametrisation of galaxies is a staple activity of many astronomers.
Indeed, it enables one to perform comparative studies and search for
correlations which hopefully provide a deeper insight into the formative and
evolutionary mechanisms at play in the Universe.
Some of the most fundamental quantities pertaining to a galaxy come from the
child-like questions: How big is it?  How bright is it?  One way astronomers
answer such apparently simple questions is through fitting model profiles to
the radial distribution of a galaxy's light.  For many years the de
Vaucouleurs (1948, 1959) $r^{1/4}$ law was employed for this task amongst
the Elliptical galaxies.  However, over the last decade or so, as the
quality of the data
has improved -- largely due to the use of CCDs -- this fitting function has
been replaced by the generalised $r^{1/n}$ profile first proposed by
S\'ersic (1968), and revitalised by Capaccioli (1987, 1989),
Davies et al.\ (1988) and Caon, Capaccioli \& D'Onofrio (1993).
In the case of Spiral galaxies,  the central bulge is also well
modelled with an $r^{1/n}$ profile (Andredakis, Peletier \& Balcells 1995;
Moriondo, Giovanardi \& Hunt 1998; Khosroshahi, Wadadekar, \&
Kembhavi 2000; Graham 2001; M\"ollenhoff \& Heidt 2001) 
-- while an exponential disk model does a remarkably good job at 
matching the observed disk light distribution.  

Although $r^{1/n}$ models fit the `observed' light profiles very closely,
there remains the question of where the galaxy actually finishes.
In practice, the sky-background often limits the extent to which one
has measured a galaxy, and the extrapolation, beyond which ever limiting
isophote this may be, is somewhat problematic.
Application of light profile models to surface brightness profiles
implicitly assumes that the galaxy profiles extend to infinity.  For a rapidly
declining intensity profile this is not such a bad assumption, as the extra
galaxy light beyond that actually observed is usually only a small percentage
of the total galaxy light.  However, in the case of the
$r^{1/4}$ law, such extrapolation can be substantial.  What this means is that
the model derived parameters of size (the effective half-light radius $r_e$)
and brightness (the surface brightness at this radius, $\mu_e$, or the mean
surface brightness enclosed by this radius, $<$$\mu$$>$$_e$), and total galaxy
magnitude, can be significantly different to those obtained in a
model-independent fashion using the `truncated' galaxy light profile.

In Section 2,
we commence by giving a little support to the $r^{1/n}$ models by
highlighting an often over-looked fact.  
The correlation between the shape parameter $n$ and galaxy size $r_e$, 
from a given sample of Elliptical galaxies, is definitely not explained 
by parameter coupling in the fitting process; this trend between 
galaxy structure and size exists when one uses model-independent values.  
We show, in Section 3, that the shape parameter $n$ is monotonically 
related with the central galaxy light concentration.  
Another issue of importance is the bias in the galaxy parameters when
one fits an $r^{1/4}$ law to a profile which is better described with
a light profile having a shape parameter $n$ which is different to 4.
In Section 4, we explore this mathematically by constructing the equations 
that govern the ratio of parameters $r_n/r_4$ (effective radii from the 
respective models) and $I_n(0)/I_4(0)$ (central intensities) when one 
forces an $r^{1/4}$ model to an $r^{1/n}$ profile.  We do
this by deriving, and solving, the analytical expressions which govern the
$\chi^2$ value which one hopes to minimise when fitting a classical 
de Vaucouleurs model to an intensity profile with shape $n$. 
These ratios are computed here as a 
function of both $n$, and the radial range $r/r_{n}$ to which one fits the
$r^{1/4}$ model.  As a result, we are able to explain why the product 
$<$$I$$>$$^\alpha_er_e$, where $\alpha$$\sim$0.7, appears 
constant, independent of whether or not one fitted an $r^{1/4}$ or an $r^{1/n}$
model -- answering a frequently mentioned question about galaxy structure.
In Section 5, 
we compute numerically the relative change to the galaxy parameters when 
one truncates the surface brightness profile at differing radii.  
We summarise our main conclusions in Section 6.

\section{The \lowercase{$n$--$\log r_e$} relation}

This section addresses the question of whether or not parameter coupling 
in the $r^{1/n}$ model can account for, or has resulted in, artificial 
correlations between the photometric parameters (see for e.g.\ Kelson 
et al.\ 2000).  One particular aspect of this potential problem is the 
correlation found between the 
S\'ersic index $n$ and the logarithm of the effective radius $r_e$ for 
Elliptical galaxies (Caon et al.\ 1993).  
If this correlation is physical, it means that the light distribution 
in Elliptical galaxies varies with galaxy size: larger galaxies tend to be 
more centrally concentrated than smaller galaxies ($n$ can be thought of as a 
central concentration parameter, see Section~\ref{GTC}). 

It is known that, irrespective of the true galaxy profile shape, in general 
the effective half-light radius derived from a fitted $r^{1/n}$ model will 
decrease and increase as the value of $n$ does.  Therefore, it is important to
verify if the trend between galaxy size and  light profile shape (that is,
structure) is physical, or simply an illusion of  the model fitting. We shall
see shortly that such a relation between structure and size  ($r_e$) {\it is}
real (that is, is  not dependent on any fitted light profile  model), and was
in fact already present, although somewhat hidden, in the  correlations between
the other global structural parameters. 

The S\'ersic $r^{1/n}$ radial intensity profile can be written as: 
\begin{equation} 
I(r)=I(0)\exp^{-b_n(\frac{r}{r_e})^{\frac{1}{n}}},
\end{equation}
where $I(0)$ is the central intensity, and $r_e$ is a scale radius.
The quantity $b_n$ is a function of the shape parameter $n$, and is chosen so 
that the scale radius encloses half of the total luminosity. 
A good approximation is $b_n$$=$$2n-0.324$ for $n$$\geq$1; however, 
in this paper we have used the exact value derived from 
$\Gamma (2n)$$=$$2\gamma (2n,b_n)$, where $\Gamma(a)$ and $\gamma(a,x)$ are 
the gamma function and the incomplete gamma function (Abramowitz \& Stegun 
1964). 

Without fitting a light profile model, the most commonly measured galaxy 
parameters are: the total galaxy luminosity $L_T$, the effective 
radius $r_e$ (defined as the equivalent radius of the isophote encircling half 
of the total galaxy flux), and the effective surface brightness $\mu_e$.  
These can be measured from the direct integration of the flux out to
some limiting isophote, and, if desired, subsequent extrapolation to 
infinity by some appropriate technique. 
These quantities are obtained without any assumption of a model (although some
ad hoc hypothesis is often done for the extrapolation to infinity), and can be 
applied to any galaxy.

\begin{figure*}
\centerline{\psfig{figure=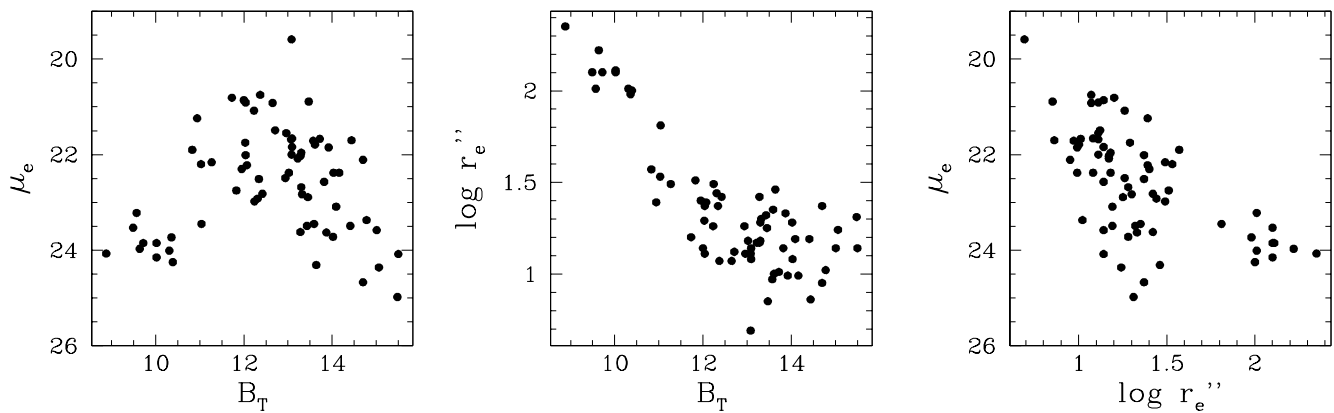,angle=-90,width=20cm}}
\caption{Correlations between the global structural parameters: total luminosity,
effective radius and effective surface brightness. Data from Caon et al.\ (1990; 1994).}
\label{Fig:3panels}
\end{figure*}

Observers have found, what are today, well known correlations between these 
photometric parameters. 
Figure~\ref{Fig:3panels} shows one example of these correlations -- based on 
the data set of Elliptical galaxies in the Virgo and Fornax Clusters (data 
from Caon et al.\ 1990, 1994). 
The existence of such correlations implies (clearly) two things: \\
a) the values are restricted to a finite region of the total parameter space, 
and \\
b) the parameters are not independent of each other.

Now, for a general photometric model, we can write the total luminosity as:

\begin{equation}
L_T = k_LI_er_e^2, \label{eq-dos} 
\end{equation}
where $k_L$ is a ``structural parameter'' whose value depends on the form of 
the galaxy light distribution (Djorgovski, de Carvalho \& Han 1988; 
Graham \& Colless 1997).  If we assume that 
the S\'ersic model can provide a good description of Elliptical galaxies, 
we can identify the $k_L$ term as 
\begin{equation}
k_L = e^{b_n}\frac{2\pi n}{b_n^{\;2n}}\Gamma(2n). \label{eq-tres}
\end{equation}
Thus, to every triplet ($L_T$, $I_e$, $r_e$) of global parameters corresponds 
a unique value of $k_L$, and hence $n$.  If all galaxies followed the 
$r^{1/4}$ law then $n$ would equal 4 for every galaxy, and $k_L$ would be 
constant for every galaxy. 

\begin{figure*}
\centerline{\psfig{figure=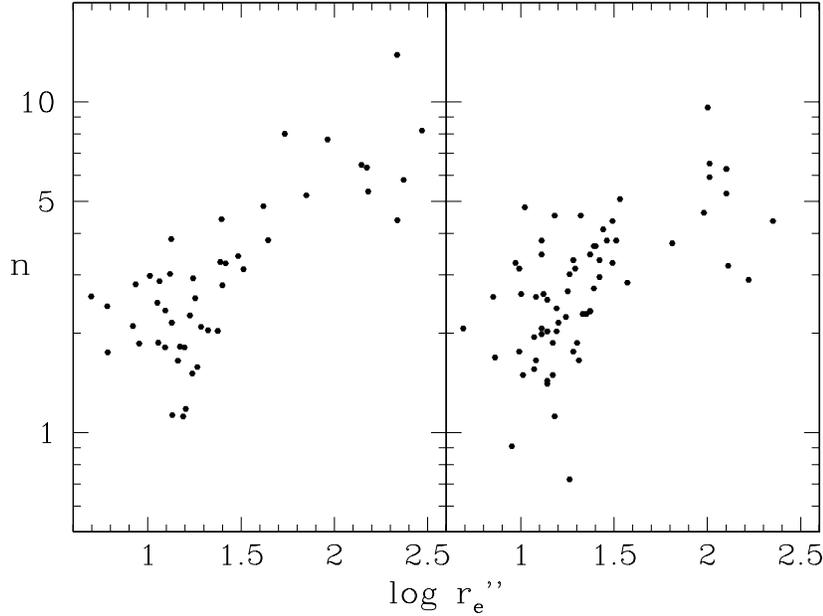,angle=-90,width=20cm,width=13cm}}

  \caption{Left panel: correlation between $n$ and $r_e$, where $n$ is derived
  from the global parameter correlations (using equations~\ref{eq-dos} and 
  ~\ref{eq-tres}); right panel: correlation between $n$ and $r_e$ using
  the values from the $r^{1/n}$ model fit to the light profile. (Data from 
  Caon et al.\ 1993).}
  \label{Fig:NparNprof}
\end{figure*}

In Figure~\ref{Fig:NparNprof}, we have plotted the values of $n$ and 
$\log r_e$.  In the left panel, the value of $r_e$ is that obtained from 
the model-independent analysis, and $n$ comes from equation~\ref{eq-tres}. 
In the right panel, both $r_e$ and $n$ were obtained from fitting a 
S\'ersic model to the surface brightness data.  The agreement is clearly 
good.  What this is telling us is that the structure of the Elliptical 
galaxies does depend on their size -- independent of any fitted photometric 
model -- one could have plotted $k_L$ instead of $n$.

To re-iterate, independent of any model, the size of an Elliptical galaxy 
is related to its structure (as given by either the little used/known 
$k_L$ term or the S\'ersic shape parameter $n$).\footnote{This correlation 
is still clear, albeit with larger scatter, if we plot $n$ against the 
old-fashioned ``effective aperture'': the radius of the circle centered on the 
nucleus within which one-half of the total flux is emitted -- as listed in the 
RC3 catalog.}  
If one chooses to characterise or model these structural shapes with the 
$r^{1/n}$ model, one finds a good agreement between the values obtained in a 
model-independent way.  In other words, the relationship between $n$ 
(or $k_L$) and $\log r_e$ is not simply due  to parameter coupling in the 
fitting routine.  
Error coupling between $n$ and $r_e$ (Graham et al.\ 1996) can ``stretch'' the 
correlation, but can not account for it entirely.\footnote{Total 
magnitudes and effective radii were measured by Caon et al.\ 
(1990, 1994) by using the $r^{1/4}$ model to extrapolate the growth 
curve to infinity.  This means that $L_T$ and $r_e$ will be 
under-estimated for profiles with $n$$>$4, and over-estimated for those 
profiles with $n$$<$4; this will shift the points in the $\log r_e$--$n$ 
diagram in the direction opposite to the parameter coupling in the S\'ersic 
model.}

When a model for the observed surface brightness profile is assumed, only
galaxies which are well described by the model should be used to obtain 
quantitative information. This can limit the number of the galaxies analysed 
in a sample. So then, why is it necessary, and/or useful, to use a model 
profile to obtain photometric parameters?   We can think of several reasons:

\begin{enumerate}
\item Although global, model-independent parameters can be obtained for
virtually any galaxy, their physical meaning can be uncertain for the 
different morphological classes, or rather, ill-defined for multiple 
component galaxies. 
\item A model provides a tool to disentangle different stellar components in
galaxies (e.g.\ S\'ersic bulge plus exponential disk).
\item By using simple analytical functions to describe the light distribution, 
it is possible to analytically obtain the density profile using the Abel 
integral equation. 
\item It allows one to obtain a more reliable description of the over-all 
structure (value of $n$) by restricting the fit  to the ``good'' part of the 
light profile, excluding, for instance, outer regions which can be affected by 
low S/N, sky-subtraction errors or distorsions due to tidal disturbances. 
\end{enumerate}

\section[]{A central concentration parameter, and its relation with \lowercase{$n$}
\label{GTC}}

Following Doi, Fukugita \& Okamura (1993), the `concentration' 
of a galaxy's light is defined in Abraham et al.\ (1994) as 
\begin{equation}
C=\frac{\sum_{i,j\in E(\alpha)}I_{ij}}{\sum_{i,j\in E(1)}I_{ij}}. 
\label{eqA1}
\end{equation}
With the radius normalised to 1 at the outer measurable isophote,
$E(\alpha)$ represents the isophote whose radius is $\alpha$ ($<$1)
times that of the outer radius of the galaxy.
$I_{ij}$ represents the intensity in the pixel ($i,j$). 

This definition is clearly ill posed when dealing with a profile model
which extends to infinity; $C$$=$$L($$<$$\alpha r)/L($$<$$r)$$\rightarrow$1 
as $r$$\rightarrow$$\infty$.   But even with a profile that is truncated
at some radius (and by this we include a radius that may encompasses all of 
the galaxy light), this definition still poses problems.  The outer 
isophote to which one reliably 
detects light is a function of exposure time, telescope aperture, 
sky-brightness, personal signal-to-noise detection requirements, etc.  
Now while a 30\% gain in radius, from a deeper 
image, may not necessarily change the total luminosity by much, it will
change the normalised radius $\alpha$ by 30\%, which can significantly 
effect the amount of light enclosed by this inner radius, and hence 
significantly effect the value of $C$. 

We therefore propose a definition for the `central' concentration which does 
not present this problem and can also be used in both a model-dependent and a 
model-independent way.  Let \TCC\ be the central concentration index, such 
that 
\begin{equation}
\TCC(\alpha)=\frac{\sum_{i,j\in E(\alpha r_e)}I_{ij}}{\sum_{i,j\in E(r_e)}I_{ij}}.
\label{eqA2}
\end{equation}
Here, $E(r_e)$ means the isophote which encloses half of the total light of 
the galaxy, and $E(\alpha r_e)$ is the isophote at a radius (0$<$$\alpha$$<$1) 
times $r_e$.  This definition is still sensitive to the outer-most radius 
used to compute the total galaxy light, but not as strongly dependent as the
definition of $C$ given in equation~\ref{eqA1}. 
For a S\'ersic law, 
\begin{equation}
\TCC(\alpha)=\frac{\gamma(2n,b_n\alpha^{1/n})}{\gamma(2n,b_n)}. 
\label{eqA3}
\end{equation}

\begin{figure*}
\centerline{\psfig{figure=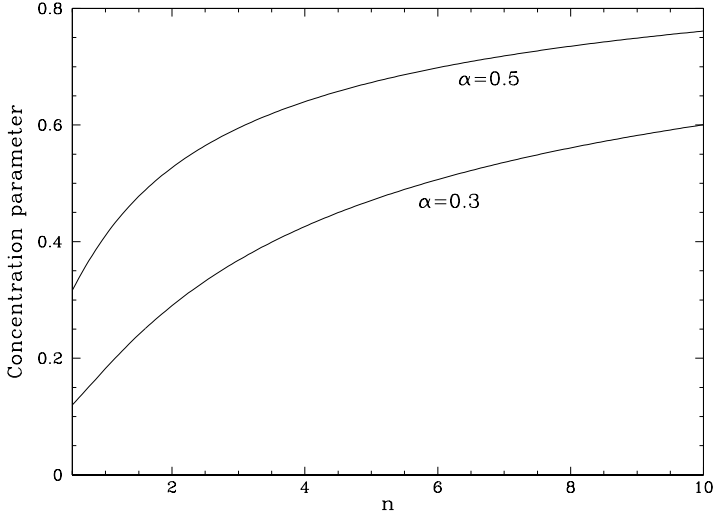,angle=-90,width=20cm,width=10cm}}
  \caption{The central concentration parameter of galaxy light, as given in 
    equation~\ref{eqA2} and ~\ref{eqA3}, is shown as a function of 
    the S\'ersic index $n$ for different values of $\alpha$.}
  \label{FigTGC}
\end{figure*}

Figure~\ref{FigTGC} shows the values of \TCC\ as a function of the S\'ersic 
index $n$, for two different values of $\alpha$, namely 0.3 and 0.5. 
The central concentration index and the index $n$ are monotonically 
related, and $n$ is therefore a useful estimator of the central concentration 
of a galaxy. 

\section[]{Why \lowercase{$r_e$}$<$$I$$>$\lowercase{$_e$}$^{0.7}$ 
is constant -- regardless of \lowercase{$n$}}

When fitting a S\'ersic model, the effective radius and mean intensity that one
derives depends on the value of $n$.  Different choices of $n$ will give
different values of $r_e$ and $<$$I$$>$$_e$ (and, of course, different $\chi
^2$ values for the fit); nevertheless, the product $r_e$$<$$I$$>$$_e^\alpha$
with $\alpha \simeq 0.7$, is extremely stable\footnote{$<$$I$$>$$_e$ means the
average surface brightness within $r_e$.  We will use the notation $r_n$ and
$<$$I$$>$$_n$ for these quantities derived from a S\'ersic profile with index
$n$. $I_n(0)$ is the central intensity at $r$$=$0.} (Kormendy \& Djorgovski
1989; Saglia, Bender, \& Dressler 1993; Kelson et al.\ 2000). As a consequence,
because the  exponent $\alpha$ almost coincides with the exponent on
$<$$I$$>$$_e$ in the Fundamental Plane scaling law (Djorgovski \& Davis 1987;
Dressler et al.\ 1987) relating $r_e$ and $<$$I$$>$$_e$ with the central
velocity dispersion $\sigma$, such that $\sigma^y$$\propto$$r_e$$<$$I$$>$$_e^x$,
the Fundamental Plane  is quite independent of whether one adopts a S\'ersic or
a de Vaucouleurs law (Kelson et al.\ 2000)\footnote{Kelson et al.\ 2000 varied
the radial extent, and hence the data, to which they fitted their models each  time they varied $n$,
making a comparison of their structural  parameters somewhat uncertain; the
influence of the sky changing  with each fitted model.}. Here, we show mathematically how the constancy of the
above product, for an individual galaxy, is a direct consequence from the form
of the S\'ersic law.

To do this we construct a profile which is perfectly described by 
a S\'ersic law with index $n$, effective radius  $r_n$ and 
central intensity $I_n(0)$. What happens if we then fit 
this profile with a de Vaucoleurs law; what values of $r_4$ and $I_4(0)$ 
shall we obtain? 

Using the $\chi^2$ goodness indicator for the fit, one can derive, 
after some work, two equations, one for each parameter, whose solutions 
provide the values of $r_4$ and $I_4(0)$ that minimise the $\chi^2$ value,
$\partial \chi^2/\partial I_4(0)$$=$0 and $\partial \chi^2/\partial r_4$$=$0. 
That is, we have derived the equations which minimise the $\chi ^2$ value 
for an $r^{1/4}$ profile from $r$$=$0 to $r$$=$$r_{\rm fin}$ (some {\it fin}al 
outer radius).

These equations are, respectively,:

\begin{equation}
4\frac{r_4}{b_4^4}\gamma\left(4,b_4\left(\frac{r_{\rm fin}}{r_n}x\right)^{1/4}\right)=
\left(\frac{I_n(0)}{I_4(0)}\right)^2S(r_{\rm fin},1) 
\label{eq-chiI}
\end{equation} 
and
\begin{equation}
4\frac{r_4}{b_4^5}\gamma\left(5,b_4\left(\frac{r_{\rm fin}}{r_n}x\right)^{1/4}\right)=
\left(\frac{I_n(0)}{I_4(0)}\right)^2S\left(r_{\rm fin},\left(\frac{r}{r_n}x\right)^{1/4}\right), 
\label{eq-chiR}
\end{equation} 
where 
\begin{equation} 
S(r_{\rm fin},f(r))\equiv\int_0^{r_{\rm fin}}e^{-2b_n\left(\frac{r}{r_n}\right)^{1/n} + b_4\left(\frac{r}{r_n}x\right)^{1/4}}f(r)dr, 
\end{equation} 
$x\equiv r_n/r_4$ and $r_{\rm fin}$ denotes the outer radius of the fitted 
profile. 
From both equations we have derived, in equation~\ref{eq-siete}, an implicit 
equation which gives the relation between $r_n$ and $r_4$ as a function of 
$n$ and $r_{\rm fin}$. 
This equation is independent of the intensity. 

\begin{equation} 
S\left(r_{\rm fin},1-b_4\frac{\gamma\left(4,b_4\left(\frac{r_{\rm fin}}{r_n}x\right)^{1/4}\right)}
{\gamma\left(5,b_4\left(\frac{r_{\rm fin}}{r_n}x\right)^{1/4}\right)}\left(\frac{r}{r_n}x\right)^{1/4}\right)=0
\label{eq-siete} 
\end{equation} 
In the limit, where $r_{\rm fin}\rightarrow\infty$, this simplifies to: 
\begin{equation} 
S\left(\infty,1-b_4\frac{\Gamma(4)}{\Gamma(5)}\left(\frac{r}{r_n}x\right)^{1/4}\right)=0
\end{equation}

\begin{figure*}
\centerline{\psfig{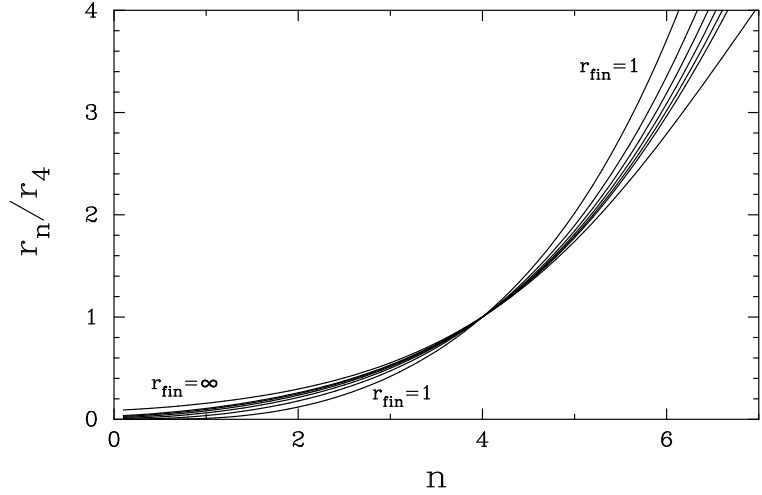}}

  \caption{The ratio of effective radii $r_n/r_4$ when one forces an $r^{1/4}$ model
  to an $r^{1/n}$ profile for values of $r_{\rm fin}$=1, 2, 3, 4, 5, 6 and $\infty$.}
\label{Fig4}
\end{figure*}

\begin{figure*}
\centerline{\psfig{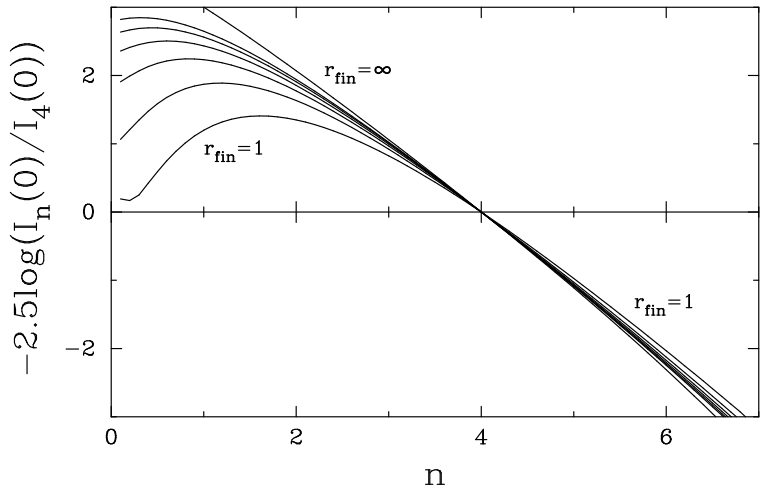}}

  \caption{The ratio of central intensities $I_n(0)/I_4(0)$ when one forces an $r^{1/4}$ model
  to an $r^{1/n}$ profile for values of $r_{\rm fin}$=1, 2, 3, 4, 5, 6 and $\infty$.}
\label{Fig5}
\end{figure*}

The solutions of this equation are shown in graphical form in 
Figure~\ref{Fig4}.  The value of $x$ can now be substituted into 
either equation ~\ref{eq-chiI} or ~\ref{eq-chiR} to obtain the ratio between 
the intensities.  
The result is shown in Figure~\ref{Fig5}.

\begin{figure*}
\centerline{\psfig{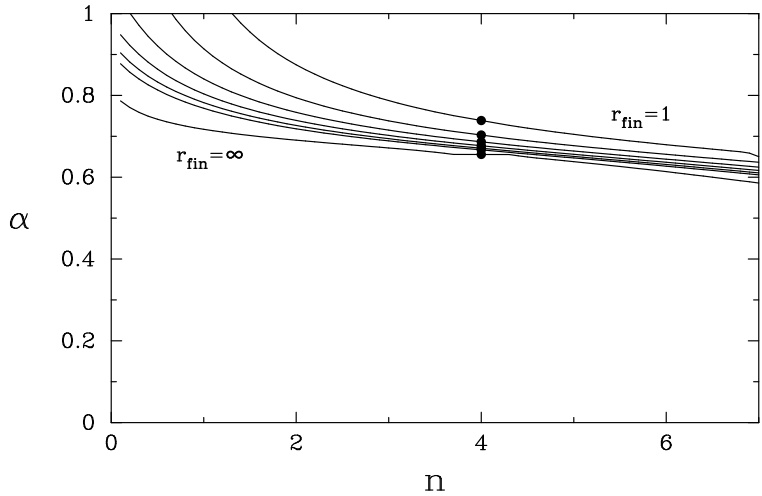}}

  \caption{The $\alpha$ exponent from equation 12 (see text for explanation) 
  for values $r_{\rm fin}$=1, 2, 3, 4, 5, 6 and $\infty$.  $\alpha$ is 
  indeterminate when $n$$=$4. }
\label{Fig6}
\end{figure*}

We can now return our attention to the exponent $\alpha$; 
from the previous results we are able to determine the value of 
$\alpha$ which solves the equation 
$\ln (r_n$$<$$I$$>$$_n^{\alpha}/r_4$$<$$I$$>$$_4^{\alpha})$$=$0, 
and thereby keeps $r_n$$<$$I$$>$$_n^{\alpha}$ roughly constant.  This is 
such that $\ln (r_n/r_4) + \alpha \ln ($$<$$I$$>$$_n/$$<$$I$$>$$_4)$$=$0, 
which can be written as 
\begin{equation} 
\alpha(n)=-\frac{\ln x}{\ln
\left[\frac{n}{4}\frac{b_4^8}{b_n^{2n}}\frac{\Gamma(2n)}
{\Gamma(8)}\left(4\frac{r_n}{x}\frac{1}{b_4^4}
\frac{\gamma(4,b_4(\frac{r_{\rm fin}}{r_n}x)^{1/4}}{S(r_{\rm fin},1)}\right)^{1/2}\right]}. 
\end{equation}

Figure~\ref{Fig6} shows the value of $\alpha$ as a function of $n$ for different 
$r_{\rm fin}$. It is noted that the value of $\alpha$ appears to be more 
or less constant at $\sim$0.7, and only weakly dependent of 
$n$ and $r_{\rm fin}$. This 
explains why the Fundamental Plane relation, as mentioned previously, 
comes out pretty much the same irrespective of which model has been used.  
In passing, we note that this is not the only relation which appears 
stable; for example, $r_e^{1/\alpha}$$<$$I$$>$$_e$ is also stable. 

While $r_e$$<$$I$$>$$_e^{0.7}$ is stable 
for individual galaxies, irrespective of the value of $n$ used in the 
S\'ersic model which derived these quantities, it seems to be a coincidence, 
rather than a natural consequence, that the same relation exists in the 
Fundamental Planes which have described the correlations between $r_e$, 
$<$$I$$>$$_e$ and central velocity dispersion for many different galaxies 
(Kelson et al.\ 2000).  
However, refined Fundamental Plane studies which have included 
the contribution from rotational energy, or used a global, rather than 
a central, velocity dispersion, have found an exponent of $\sim$1, 
in agreement with the expectation from the virial theorem 
(Busarello et al.\ 1997; Graham \& Colless 1997; Prugniel \& Simien 1997). 
We would therefore argue that for 'refined' Fundamental Plane studies, 
one should be concerned about the real range of structural shapes (as 
evidenced in Figure~\ref{Fig:NparNprof}), and not be contempt with a model
that ignores this; $r_e$$<$$I$$>$$_e^{1.0}$ is not invariable with $n$.

\section[]{Model-dependent versus model-independent parametrisation}

Although the photometric parameters $L_T$, $r_e$ and $<$$\mu$$>$$_e$ are, at
least mathematically, well defined quantities, their values do depend on the 
method used to measure them.
Not just with regard to the particular form of the fitting function 
(i.e.\ the model used to fit either the surface brightness profile, 
or the `curve of growth'), but also with regard to the radial extent 
to which the galaxy profile is assumed to hold. 

The curve of growth can asymptote into the noise well before the galaxy 
actually peeters out -- truncated by the sky-background and short exposure 
times. Summing the galaxy magnitude within the isophotal ellipse where this 
occurs can result in an under-estimation of the total galaxy magnitude, and 
hence an under-estimation to the effective half-light radius, 
and an over-estimate to intensity of the surface brightness term.  
On the other hand, assuming that the galaxy surface brightness profile follows,
all the way to infinity, the best-fitting model to the inner data points, may 
result in the attribution of substantially more light to the galaxy than 
actually exists.

\subsection{Total galaxy luminosity}

The effective half-light radius derived from the S\'ersic model will be
denoted $r_{e,{\rm mod}}$ from here on, to avoid confusion with the similar 
term derived in a model-independent way. 
The total luminosity associated with an $r^{1/n}$ model profile can be 
written as 
\begin{equation}
L_T=I(0)r_{e,{\rm mod}}^2\frac{2\pi n}{b_n^{2n}}\Gamma(2n). 
\label{lum1}
\end{equation}

When using the integrated surface brightness profile, or growth curve, the 
observer selects a finite radius, $r_{\rm fin}$, where the curve of growth 
becomes flat. The value of $r_{\rm fin}$ which one selects depends on the 
exposure time, and hence noise, in the outer parts of the galaxy image.  
If one accepts that the surface brightness profile out to $r_{\rm fin}$ is 
well described by a S\'ersic law, contributing zero light at larger radii, 
then the `total' luminosity obtained from direct measurements of the 
integrated surface brightness profile is
 
\begin{equation}
L(r_{\rm fin})=I(0)r_{e,{\rm mod}}^2\frac{2\pi n}{b_n^{2n}}
\gamma\left(2n,b_n\left(\frac{r_{\rm fin}}{r_{e,{\rm mod}}}\right)^\frac{1}{n}\right). 
\label{lum2}
\end{equation}

For an $r^{1/n}$ profile that extends to infinity, the outer fraction of the 
total galaxy light beyond the radius $r_{\rm fin}$ is given by 
\begin{equation}
F(r_{\rm fin})\equiv\frac{L_T-L(r_{\rm fin})}{L_T}=1-
\frac{\gamma(2n,b_n(\frac{r_{\rm fin}}{r_{e,{\rm mod}}})^\frac{1}{n})}{\Gamma(2n)}
\end{equation}

\begin{figure*}
\centerline{\psfig{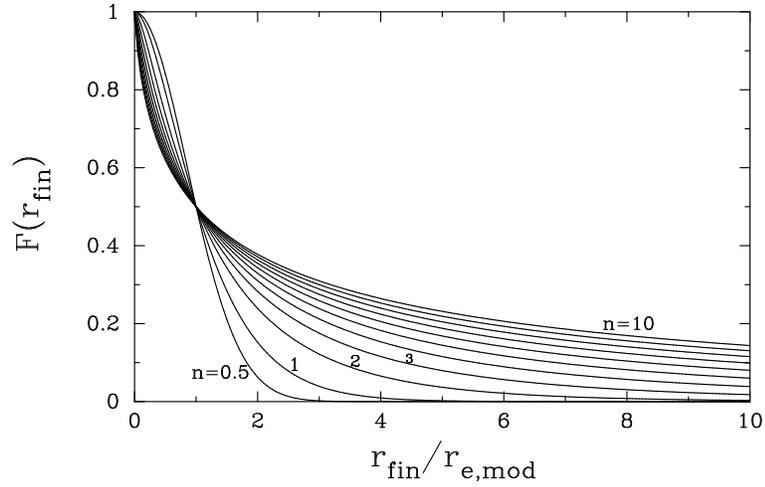}}

  \caption{The parameter $F(r_{\rm fin})$ is shown as a function of 
   $r_{\rm fin}/r_{e,{\rm mod}}$ for values of n=0.5,1,2,3,...,10.}
  \label{Fig7}
\end{figure*}

This fractional difference to the total luminosity is plotted in 
Figure~\ref{Fig7}, as a function of $r_{\rm fin}/r_{e,{\rm mod}}$, for different 
values of $n$.  
When $r_{\rm fin}/r_{e,{\rm mod}}$$=$1, the fraction of the total galaxy 
luminosity outside $r_{\rm fin}$ is, by definition, 50 per cent (or 0.75 mag). 
For the de Vaucouleurs profile ($n$$=$4) the outer fraction to the luminosity 
is $\sim$15 per cent (0.18 mag) at a radius $r_{\rm fin}$ equal to four 
$r_{e,{\rm mod}}$.

\begin{figure*}
\centerline{\psfig{figure=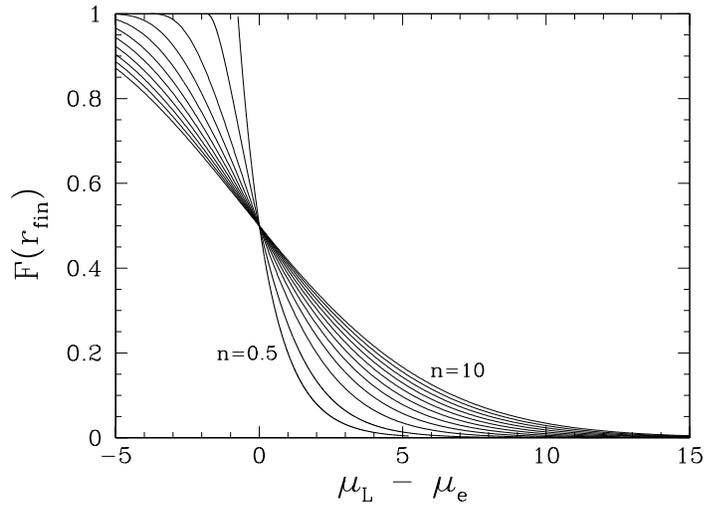,angle=-90,width=20cm,width=10cm}}

  \caption{The parameter $F(r_{\rm fin})$ is shown as a function of
   $\mu_{\rm L}-\mu_e$ for values of $n=0.5$,1,2,3,...,10.}
  \label{Fig8} 
\end{figure*}

Usually an observer knows, either in advance when preparing the observations,
or after extraction of the light profiles, the surface brightness limit 
$\mu_{\rm L}$ of the image (that is, the surface brightness level of
the last measurable isophote). 
For this reason, it is useful to plot (Figure~\ref{Fig8}) the parameter 
$F(r_{\rm fin})$ against the difference $\mu_{\rm L}-\mu_e$ for 
various values of $n$.  Here, $F(r_{\rm fin})$ is such that 
\begin{equation}
F(r_{\rm fin})=1-
\frac{\gamma(2n,b_n+\frac{\ln(10)}{2.5}(\mu_L-\mu_e))}{\Gamma(2n)}
\end{equation}

It is evident that, to reach the same value of $F(r_{\rm fin})$, the
difference $\mu_{\rm L}-\mu_e$ must be larger (that is, the images deeper)
for bigger $n$. This is particularly important because $\mu_e$ and $n$ are 
positively (though weakly) correlated.
For instance, the $\mu_e$--$n$ plot from Caon et al.\ (1993) 
shows that, when $n=2$, $\mu_e$$\simeq$22.5, while at $n$$=$8 
$\mu_e$$\simeq$23.5 (in B). 
The growth curve integrated out to a limiting surface brightness of 
$\mu_{\rm L}$$=$27 mag/arcsec$^2$ will miss about 5\% of the total 
luminosity when $n$$=$2, and 22\% when $n$$=$8.

\subsection{Effective radius}

The two different estimates to the total galaxy luminosity (Eq. \ref{lum1} and 
Eq. \ref{lum2}) result in two different estimates for the effective half-light 
radius. 
Taking the truncated profile, we define the observed effective half light 
radius $r_{e,{\rm obs}}$ such that 

\begin{equation}
L(r_{e,{\rm obs}})\equiv\frac{L(r_{\rm fin})}{2}. 
\end{equation}
This can be expanded to give 
\begin{equation}
2\gamma(2n,b_n(\frac{r_{e,{\rm obs}}}{r_{e,{\rm mod}}})^\frac{1}{n})=
\gamma(2n,b_n(\frac{r_{\rm fin}}{r_{e,{\rm mod}}})^\frac{1}{n}).
\end{equation}
 
\begin{figure*}
\centerline{\psfig{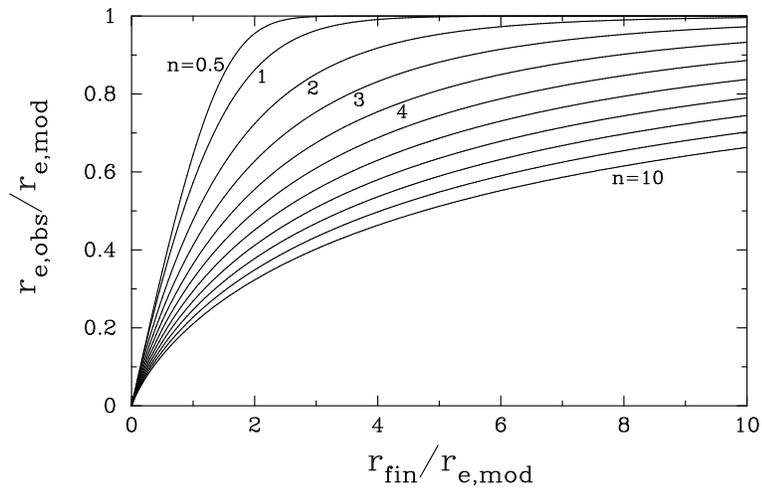}}

  \caption{The $r_{e,{\rm obs}}/r_{e,{\rm mod}}$ ratio as a function of 
	$r_{\rm fin}/r_{e,{\rm mod}}$ for different values of $n$ 
        (see the text for an explanation of terms). }
  \label{Fig9}
\end{figure*}

Once the value of $r_{\rm fin}$ is selected, and $r_{e,{\rm mod}}$ and $n$ 
derived, one can compute the ratio between $r_{e,{\rm obs}}$ and 
$r_{e,{\rm mod}}$.  Figure~\ref{Fig9} shows this ratio for different values 
of $n$ and $r_{\rm fin}/r_{e,{\rm mod}}$.  
As $n$ increases, for a fixed $r_{\rm fin}/r_{e,{\rm mod}}$ ratio, the 
$r_{e,{\rm obs}}/r_{e,{\rm mod}}$ ratio decreases from 1.
For a de Vaucouleurs profile observed out to 4$r_{e,{\rm mod}}$, the radius 
containing half of the `observed' galaxy light, $r_{e,{\rm obs}}$, is 3/4 of 
the radius, $r_{e,{\rm mod}}$, coming from the model extrapolation to infinity.

\subsection{Effective surface brightness}

The S\'ersic surface brightness profile $\mu (r)$ can be written as 
\begin{equation}
\mu (r)=\mu _{0}+\frac{2.5b_n}{\ln(10)}\left( \frac{r}{r_{e,{\rm mod}}}\right) ^{1/n},
\end{equation}
where $\mu _{0}$ is the central surface brightness. 
The effective surface brightness, $\mu_e$, is the surface brightness at the 
effective half-light radius.  Therefore, for an $r^{1/n}$ model which extends
to infinity, it is given by 
$\mu_{e,{\rm mod}}$$=$$\mu _{0}$$+$$2.5b_n/\ln(10)$. 
The difference between the model value and the value assuming a truncated
profile is 
\begin{equation}
\Delta \mu_e  = \frac{2.5b_n}{ln(10)}\left[ 1 - \left( \frac{r_{e,{\rm obs}}}{r_{e,{\rm mod}}}\right)^{1/n} \right]. 
\end{equation}

\begin{figure*}
\centerline{\psfig{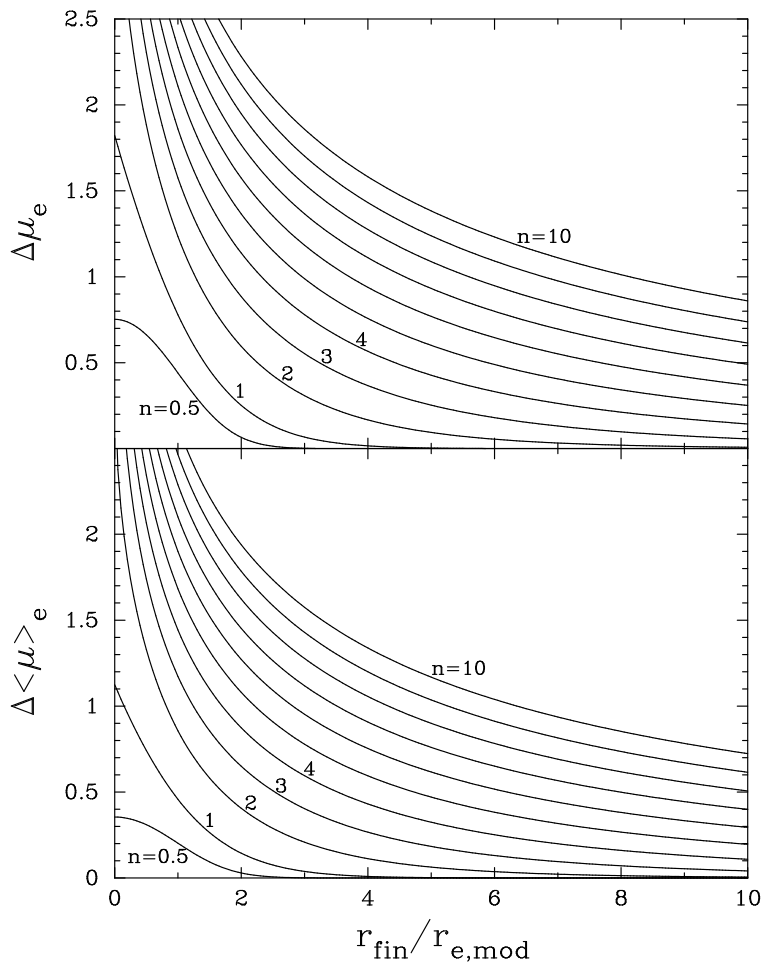}}

  \caption{The difference in the effective surface brightness $\mu_e$ (upper
  panel) and mean effective surface brightness $<$$\mu$$>$$_{e}$
   as a function of $r_{\rm fin}/r_{e,{\rm mod}}$ for different values of $n$.}
  \label{Fig10}
\end{figure*}

The mean effective surface brightness is defined as:
\begin{equation}
<\mu>_e\equiv-2.5\log\frac{L(r_e)}{\pi r_e^2}, 
\end{equation}
where $r_e$ can be either $r_{e,{\rm mod}}$ or $r_{e,{\rm obs}}$. 
The difference, 
$\Delta$$<$$\mu$$>$$_{e}$$\equiv$$<$$\mu$$>$$_{e,{\rm mod}}$$-$$<$$\mu$$>$$_{e,{\rm obs}}$
is given by the expression 
\begin{equation}
\Delta<\mu>_e=-2.5\log\left[
\frac{\Gamma(2n)}{2\gamma(2n,b_n(\frac{r_{e,{\rm obs}}}{r_{e,{\rm mod}}})^\frac{1}{n})}  
\left(\frac{r_{e,{\rm obs}}}{r_{e,{\rm mod}}}\right)^2\right]
\end{equation}

Figure~\ref{Fig10} shows both $\Delta \mu_e$ and $\Delta$$<$$\mu$$>$$_{e}$ as a 
function of $r_{\rm fin}/r_{e,{\rm mod}}$ for different values of $n$.

\section[]{Conclusions}

We have shown that the correlation between Elliptical galaxy size ($r_e$)
and luminous structure ($n$, or $k_L$) is not only real, but was 
already contained in the known correlations between the other global, 
model-independent photometric parameters: $L_T$, $r_e$ and $I_e$. 
While parameter coupling in the fitting process can certainly contribute 
to the observed correlation, it can not account for it.

We have redefined the galaxy concentration index, rendering it almost 
independent of the limiting magnitude and/or radius of the galaxian map.  
For a S\'ersic model, the new index displays a monotonic behaviour with $n$.

We have shown, through a mathematical analysis, why the quantity 
$<$$I$$>$$^\alpha_er_e$, with $\alpha$$\sim$0.7, is fairly constant and 
insensitive to both $n$ and the radius out to which the profile 
is fitted.  Intriguingly, because $\alpha$ is practically equal to the 
exponent on the intensity term in the Fundamental Planes constructed 
with the quantities $r_e$, $<$$I$$>$$_e$ and central velocity dispersion, 
these Fundamental Planes are insensitive to galaxy structure.  However, 
more refined studies which further take into account dynamical non-homology, 
rotational energy and/or consider metallicity effects, find a different 
exponent.  Consequently, the use of an $r^{1/4}$ law to obtain $r_e$ and 
$<$$I$$>$ can affect these Fundamental Plane relations. 

Galaxian images are generally limited in their radial extension by noise,
sky subtraction errors or distorsions in the outer parts; it is therefore
useful to know how the photometric parameters vary as a function of the
radius out to which the surface brightness distribution is integrated.
Galaxies described with larger shape parameters ($n$) require more extended
(in units of $r_e$), or deeper (with respect to $\mu_e$), observations in
order not to miss an important fraction of the total light.  For example, when
$n\geq4$ one will obtain significantly smaller effective radii (by a factor of 
2 or more) and brighter effective surface brightnesses (by 1 mag or more) if 
one only integrates to $\sim$2 effective radii. 

Unfortunately, the radial extent to galaxian images cannot 
simply be increased as far as one desires by integrating for 
longer, even if galaxies are of infinite radius. 
If the limiting factor were the photon noise from a homogeneous sky 
background, then the signal-to-noise ratio would be proportional to 
the square root of the integration time.   However, a further 
contribution to the sky background noise is given by scattered 
light and by faint sources which make up the extragalactic 
background light (a vivid representation is provided by imagining 
a galaxy superposed on the Hubble Deep Field). 
Dalcanton and Bernstein (2000) carried out a full analysis of the sky 
background noise sources and of the limiting surface brightness which can 
be achieved by deep imaging, finding limits of $\mu_B \sim 29.5$ mag 
arcsec$^{-2}$ and $\mu_R \sim 29$ mag arcsec$^{-2}$.
This corresponds to $\sim 10$ effective radii for a bright elliptical with 
$\mu_e(B) \simeq 23.5$ and $n=8$, and $\sim 7$ effective radii for a 
low-luminosity elliptical with $\mu_e(B) \simeq 22.5$ and $n=2$.

In Caon et al.\ (1990, 1994) the light profiles of elliptical galaxies, 
reach a surface brightness limit of $\mu_B \simeq 27.5$ (typically 
4--5 effective radii).  There is no evidence for an outer truncation 
similar to that observed in the disks of spirals (Pohlen, Dettmar \& 
{\"u}tticke 2000 and references therein).  As far as we know, nobody so 
far has detected definite edges in elliptical galaxies.

\section*{Acknowledgments}
We are indebted to Ignacio Garc\'{\i}a de la Rosa and Aurora Sicilia who 
helped initiate the early phases to some of the issues addressed here.

\bsp

\label{lastpage}

\end{document}